\documentstyle
[12pt]{article} \hoffset -0.5in \textwidth 6.5in \textheight
9.00in  \parskip 7pt \openup2.5\jot \parindent=0.5in
\renewcommand{\thefootnote}{\fnsymbol{footnote}}

\def\lsim{\;\centeron{\raise.35ex\hbox{$<$}}{\lower.65ex\hbox
{$\sim$}}\;}
\def\gsim{\;\centeron{\raise.35ex\hbox{$>$}}{\lower.65ex\hbox
{$\sim$}}\;}
\def\centeron#1#2{{\setbox0=\hbox{#1}\setbox1=\hbox{#2}\ifdim
\wd1>\wd0\kern.5\wd1\kern-.5\wd0\fi
\copy0\kern-.5\wd0\kern-.5\wd1\copy1\ifdim\wd0>\wd1
\kern.5\wd0\kern-.5\wd1\fi}}

\newskip\humongous \humongous=0pt plus 1000pt minus 1000pt

\newif\ifdtup

\relax
\def\np#1#2#3 {Nucl. Phys. \underline{#1} (19#2) #3}
\def\prl#1#2#3 {Phys. Rev. Lett. \underline{#1} (19#2) #3}
\def\prev#1#2#3 {Phys. Rev. \underline{#1} (19#2) #3}
\def\pl#1#2#3 {Phys. Lett. \underline{#1} (19#2) #3}
\def\rmp#1#2#3 {Rev. Mod. Phys. \underline{#1} (19#2) #3}
\def\prep#1#2#3 {Phys. Rep. \underline{#1} (19#2) #3}
\def\spu#1#2#3 {Sov. Phys.-Usp. \underline{#1} (19#2) #3}
\def\sjnp#1#2#3 {Sov. J. Nucl. Phys. \underline{#1} (19#2) #3}
\def\jetp#1#2#3 {JETP Lett. \underline{#1} (19#2) #3}
\def\apj#1#2#3 {Astrophys. J. \underline{#1} (19#2) #3}
\def\apjl#1#2#3 {Astrophys. J. Lett. \underline{#1} (19#2) #3}
\def\ib#1#2#3 {{\it ibid.} \underline{#1} (19#2) #3}
\def\nat#1#2#3 {Nature (London) \underline{#1} (19#2) #3}
\def\ap#1#2#3 {Ann. Phys. (NY) \underline{#1} (19#2) #3}
\def\nc#1#2#3 {Nuovo Cim. \underline{#1} (19#2) #3}
\def\zp#1#2#3 {Zeit. Phys. \underline{#1} (19#2) #3}
\def\ar#1#2#3 {Ann. Rev. Nucl. Part. Sci. \underline{#1} (19#2) #3}
\def\prs#1#2#3 {Proc. Roy. Soc. \underline{#1} (19#2) #3}
\def\pcps#1#2#3 {Proc. Cam. Phil. Soc. \underline{#1} (#2) #3}
\def\rpp#1#2#3 {Rep. Prog. Phys. \underline{#1} (19#2) #3}
\def\cpc#1#2#3 {Computer Phys. Comm. \underline{#1} (19#2) #3}
\def\ptp#1#2#3 {Prog. Th. Phys. \underline{#1} (19#2) #3}
\def\ijmp#1#2#3 {Int. J. Mod. Phys. \underline{#1} (19#2) #3}
\def\app#1#2#3 {Acta. Phys. Pol. \underline{#1} (19#2) #3}
\def\mpl#1#2#3 {Mod. Phys. Lett. \underline{#1} (19#2) #3}
\def\beq{\begin{equation}}
\def\eeq{\end{equation}}

\def\c{\gamma}   \def\C{\Gamma}
\def\d{\delta}   
\def\e{\epsilon}

\def\g{\eta}
\def\h{\theta}

\def\m{\mu}
\def\n{\nu}
\def\p{\pi}

\def\VEV#1{\left\langle #1\right\rangle}

\def\aem{\alpha_{\mbox{\scriptsize EM}}}
\relax
\begin{document} \begin{titlepage}
\rightline{\vbox{\halign{&#\hfil\cr
&ANL-HEP-PR-93-4\cr
&BROWN-HET-872\cr
&\today\cr}}}
\vspace{0.25in}
\begin{center}

{\Large\bf
THE $\eta_6$ AT LEP AND TRISTAN}\footnote{Work
supported by the U.S. Department of
Energy, Division of High Energy Physics, Contract\newline W-31-109-ENG-38 and
contract DE-FG02-91ER40688-TaskA.}
\medskip

\normalsize  Kyungsik Kang
\\ \smallskip
Physics Department\\Brown University
\\Providence, RI 02912

Ian G. Knowles and Alan R. White
\\ \smallskip
High Energy Physics Division\\Argonne National
Laboratory\\Argonne, IL 60439\\ \end{center}

\begin{abstract}
      The $\eta_6$ is a ``heavy axion'' remnant of dynamical electroweak
symmetry breaking by a color sextet quark condensate. Electroweak scale color
instanton interactions allow it to be both very massive and yet be
responsible for Strong $CP$ conservation in the color triplet quark
sector. It may have been seen at LEP via its two-photon decay mode and
at TRISTAN via its hadronic decay modes.

\end{abstract}

\renewcommand{\thefootnote}{\arabic{footnote}} \end{titlepage}

Electroweak dynamical symmetry breaking by a chiral condensate of color sextet
quarks\cite{sex} has many theoretically attractive features, including the
special resolution of the Strong $CP$ problem via a heavy axion that we
outline below. However, it should also be emphasised that if this should
turn out to be the path chosen by nature it provides a particularly inviting
prospect for experimental high-energy physics. Because of the direct
coupling of the strong and electroweak interaction, the spectrum of new
phenomena that can be expected to appear, at both currently operating
accelerators and the future $LHC$ and $SSC$ machines, is probably at least as
large, if not considerably larger, than in any other symmetry-breaking
scenario.

The purpose of this paper is to focus on the tantalizing possibility
that a distinctive feature of the symmetry breaking, namely the
``heavy axion'' $\eta_6$, has already been seen experimentally. This
particle is expected to have both a major two-photon decay mode and
characteristic high multiplicity hadron decays. It is therefore an excellent
candidate for the new particle, with a mass of 59 GeV, suggested by the
two-photon pairs seen at LEP\cite{lep}. That a small bump is also
seen\cite{trist} at TRISTAN, at just this energy, can then be interpreted as
due to its hadronic decay modes.

We expect the strong interaction dynamics of the sextet quark sector of
$QCD$ to be very different from that of the triplet sector. In
particular we anticipate that relatively complicated instanton generated
interactions (at and above the electroweak scale), which include
``Strong'' $CP$-violating effects, will play an important
role\cite{arw,hol}. As a
result, only a minimal amount of rescaling of physics from the triplet
to the sextet sector will be possible. We will (if candidate sextet
phenomena begin to appear) be studying a new realm of gauge theory
physics and it will not be surprising if, to a large extent, the theory
has to follow along semi-phenomenologically behind the experimental
discoveries.

Although it is correctly described as a heavy axion, the $\eta_6$ is the
``Higgs particle'' of sextet symmetry breaking in the sense that its
experimental discovery would be the most immediate confirmation of this
form of symmetry breaking. Motivated, in part, by the anomalous real part
measured in elastic $\bar{p}p$ scattering at the apparent threshold
energy\cite{UA4}, we suggested\cite{ka} that the $\eta_6$ be identified
with a very heavy state, with a mass\cite{ari} of O(60) GeV, seen in exotic
Cosmic ray events\cite{has}. Because of the axion nature of the $\eta_6$, we
proposed that this particle be looked for in accelerators via its two-photon
decay mode.

As is by now well known\cite{lep}, the L3, DELPHI, and ALEPH experiments
at LEP have recently reported several events of the form
$Z^0 \to l^+l^- + \gamma\gamma$, in which the mass of the $\gamma\gamma$ pair
is O(60) GeV. The lepton pairs are either muons or electrons and we have
separately plotted the $m_{\gamma\gamma}$ distribution for muon and
electron pairs in Fig.~1. There are as yet, no neutrino or $\tau$ pairs,
although DELPHI has one candidate quark pair event. While the kinematics
of some of the events may be compatible with $QED$ radiation, others
look implausible explained this way. In particular the muon events in
the 59 GeV bin are all ``large angle'' events and do not look like naive
radiative events. Rather they suggest the existence of a new
``particle'', i.e. resonance, with a mass of 59 GeV and a width
of (up to) O(1) GeV. Clearly a case could be made from Fig.~1
that {\em only the muon pair events} suggest a new resonance.
(Particularly since the two electron events close to 59 GeV are both
good candidates for QED radiation.) This is potentially a significant feature,
as we shall see.

Since we expect the $\eta_6$ to have hadronic decay modes involving
relatively complicated high multiplicity states, it is particularly interesting
that the new ``particle'' may also have been seen at TRISTAN. In fact all three
experiments saw\cite{trist} a small peak in the hadronic cross-section at
59.05 GeV. An error-weighted average of the TRISTAN results for R is
shown in Fig.~2. AMY actually obtained a value more than
$30\%$ above the standard model value (although with a large error -
giving at most a ``2$\sigma$ effect''). If this effect is produced by
the same new particle that appears in the LEP events, we can infer both
that it couples to electrons and that it does indeed have {\it major
hadronic decay modes}. If this particle had direct electroweak
couplings to quark and lepton states in analogy, say, with the $Z^0$,
then the corresponding decays would surely have already been seen at LEP. It
seems more likely to us that the width is produced mostly by the photon
pairs and high multiplicity hadron states which would not be so easily
identified at LEP, but clearly would be registered at TRISTAN.

We can briefly summarise the essentials of sextet symmetry breaking\cite{sex}
as follows. A massless flavor doublet $(U,D)$ of color sextet quarks with the
usual quark quantum numbers (except that the role of quarks and antiquarks is
interchanged) is first added to the Standard Model {\it with no scalar
Higgs sector}. Within $QCD$, conventional chiral dynamics will
break the sextet axial flavor symmetries spontaneously and produce four
massless pseudoscalar mesons (Goldstone bosons), which we denote as
$\pi^+_6,\;\pi^-_6,\;\pi^0_6$ and $\eta_6$. The $\pi^+_6,\;\pi^-_6$, and
$\pi^0_6$  are ``eaten'' by the massless electroweak gauge bosons and
respectively become the third components of the massive $W^+,\;W^-$ and
$Z^0$ - giving $M_W\sim g\,F_{\pi_6}$ where $F_{\pi_6}$ is {\it a $QCD$ scale}.
$F_{\pi_6} \sim 250 GeV$ is consistent with an elementary ``Casimir Scaling''
rule\cite{sex}.

The $\eta_6$ is not involved in generating mass for the electroweak gauge
bosons and remains a Goldstone boson. A first assumption might be that the
$\eta_6$ somehow acquires an electroweak scale mass which is nevertheless
small enough that we can utilise $PCAC$ for the sextet $U(1)$ axial
current. The analog of the familiar $\pi^0 \to 2\gamma$ calculation, but
involving the sextet quark triangle anomaly, will give amplitudes for
\beq
\label{amp}
\eta_6 \to \gamma\gamma, ~~~Z^0 \to \eta_6 + \gamma~~~~ \mbox{and}
{}~~ Z^0 \to \eta_6 + Z^{0\star}.
\eeq
where the $Z^{0\star}$ is an off-shell $Z^0$.

Note that if the $\eta_6$ is a pseudoscalar, and {\em $CP$ is conserved}, the
existence of just two independent momenta implies that each of the vertices
in (\ref{amp}) must have the ``pseudotensor'' kinematic form
\beq
\label{vertep}
\C_{\m\n}=C(p,q)\e_{\m\n\c\d}p^\c q^\d
\eeq
where $p$ and $q$ are the momenta involved, and $C(p,q)$
can be calculated from the anomaly. Assuming a mass of 60 GeV,
gives\cite{ros} a very narrow width of 0.17 keV for $\eta_6 \to \gamma\gamma$.
For $Z^0 \to \eta_6 + \gamma$ the anomaly calculation\cite{hu} predicts one
event in 20 million at LEP, while from the $Z^0 \to \eta_6 + Z^{0\star}$
calculation we obtain a rate for $Z^0 \to \eta_6 + Z^{0\star} \to \eta_6 +
\mu^+\mu^-$ of 2 events in a billion. This is at least three orders of
magnitude too small to explain the two photon events. We conclude that,
in general, the anomaly gives a set of amplitudes which are far too
small to be compatible with the LEP events.

The sextet quark anomaly estimates for amplitudes can only be significantly
wrong if Goldstone boson intermediate states can contribute to the processes
involved. At this point the special ``heavy axion'' nature of the $\eta_6$
becomes crucial. Apart from its high mass, the $\eta_6$ is actually a
``Peccei-Quinn axion'' and can be responsible for Strong $CP$ conservation
{\em in the triplet quark sector} in a conventional manner\cite{pec}. However,
if the $\eta_6$ is the origin of $CP$ conservation in the triplet
sector, then {\em the sextet sector} (and sextet Goldstone boson amplitudes
in particular) {\em will not be $CP$ conserving}. As a
result, there will be intermediate states contributing to ``longitudinal''
$Z^0$ and $W^{\pm}$ amplitudes involving the $\eta_6$ which do
invalidate the anomaly estimates, and could give large enough
cross-sections at LEP and TRISTAN. We can briefly summarise the physics
behind the $CP$-related properties of the $\eta_6$ as follows.

The Peccei-Quinn argument for Strong $CP$ conservation requires\cite{pec} the
existence of a Goldstone Boson axion $a$ that couples to the $QCD$ color
anomaly and gives an effective lagrangian {\em for the triplet quark sector}
of the form
\beq
\label{laga}
{\cal L}={\cal L}_{\mbox{\scriptsize QCD}}+\tilde{\h}\frac{g^2}{32\p^2}
F\tilde{F}+\frac{a}{\mbox{v}_{\mbox{\scriptsize PQ}}}\frac{g^2}{32\p^2}
F\tilde{F}+\cdots
\eeq
where ${\cal L}_{\mbox{\scriptsize QCD}}$ is the usual QCD lagrangian for
the gauge and triplet quark sectors and, in a conventional notation,
$\tilde{\h}=\h+\arg\det m_3$, where $m_3$ is the triplet quark mass matrix.
v$_{\mbox{\scriptsize PQ}}$ is the vacuum condensate which produces the
Goldstone Boson axion $a$. An appropriate shift in $a$ will absorb the
$CP$-violating $\tilde{\h}$ term and a sufficient condition for the minimum
of the axion potential to occur at $\hat{\h}=0$ (where now
$\hat{\h}=\h+\arg\det m_3+\VEV{a}/\mbox{v}_{\mbox{\scriptsize PQ}}$) is that
$\VEV{F\tilde{F}}$ vanishes like $\sin\hat{\h}$ at $\hat{\h}=0$. This
is expected to be the case for normal instanton interactions. A mass for
the axion is generated by the curvature of the potential at the minimum.
If all of the relevant $QCD$ dynamics involves only the normal $QCD$
scale $\Lambda_{\mbox{\scriptsize QCD}}$, this mass is inevitably of
O($\Lambda^2_{\mbox{\scriptsize QCD}}/\mbox{v}_{\mbox{\scriptsize PQ}}$) and
hence very small\cite{pec}.

If we identify $a$ with the $\eta_6$, the mass can be much higher
just because of the intricate $QCD$ dynamics at the sextet scale. To
generate the usual quark and lepton masses it is necessary to add
four-fermion couplings to the theory which combine
appropriately with the $\VEV{\bar{Q}Q}$ sextet condensate . If we then obtain
(\ref{laga}) by integrating out the sextet quark
sector, we must include $\g_6$ vertices induced\cite{hol} by the combination of
$\bar{Q}Q\bar{q}q$ vertices, the $\VEV{\bar{Q}Q}$ condensate, and
instanton interactions involving sextet quarks. The instanton interactions are
actually {\em very high order fermion vertices}. The simplest such
vertex involves each flavor of triplet quark (and antiquark) and {\em
five} of each flavor of sextet quark (and anti-quark). When the
condensate and four-fermion vertices are
combined with the instanton vertices, a large array of
interactions is obtained. (Indeed the resulting low order vertices may
be enhanced by large factorial factors associated with the possibilities
for pair condensation.) These fermion vertices can then be coupled by
arbitrarily complicated gluon interactions - which are effectively
infra-red interactions at the sextet scale.

Note that with the additional two flavors of sextet quarks, the
resulting evolution of $\alpha_s$ is negligible above the electroweak
(sextet) scale. Indeed there is an effective {\em infra-red} fixed-point
controlling the dynamics of the sextet $QCD$ sector\cite{arw}. The associated
absence of the infra-red growth of the gauge coupling implies that, in
this sector, confinement and chiral symmetry breaking will involve the
instanton interactions we are discussing as an important ``infra-red''
effect. (There are no infra-red renormalons\cite{arw} and so instantons
don't melt!). For our present purposes all that we need extract from
this complicated dynamical situation is that the sextet instanton
interactions generating $\eta_6$ vertices all contain a factor\cite{hol}
of $\cos [\tilde{\theta}+<\eta_6>]$.  Therefore an axion potential of
the form $V(\cos~[\tilde{\theta}~+~<\eta_6>])$ is generated. Such a
potential will naturally retain the $CP$-conserving minimum at
$~\tilde{\theta}+<\eta_6>=0$ while also giving an $\eta_6$ mass (the
curvature at the minimum) of order the electroweak scale - say 60 GeV!

Focussing now on the $CP$ properties of the sextet sector, we note that
the Peccei-Quinn argument is inapplicable since we can
not write a lagrangian of the form (\ref{laga}) - that is involving both the
$\eta_6$ and the gluon field - to describe sextet quark interactions. If the
gluon field is to be present, then we must use the full $QCD$
lagrangian, written in terms of elementary fields, for the combined
triplet and sextet sectors. This clearly has no axion. Also, we know
that the four-fermion $\bar{Q}Q\bar{q}q$ couplings that we add to the
theory must be $CP$-violating since they have to
produce the $CP$-violating triplet quark
mass matrix. Because there is {\em no axion}, the induced fermion
vertices involving instanton interactions will automatically be $CP$
(and separately $C$) -violating. In effect, a consequence of the usual
$CP$ violation in the triplet quark masses is that the ``low-energy'' effective
lagrangian for $QCD$ interactions of the $\eta_6,\;\pi^+_6,\;\pi^-_6$,
and $\pi^0_6$ is necessarily $CP$-violating. In unitary gauge, it is the
``longitudinal'' (or scalar) components of the gauge boson fields, i.e.
$\partial^{\mu}Z^0_{\mu},\;\partial^{\mu}W^+_{\mu}$ and
$\partial^{\mu}W^-_{\mu}$, that inherit the interactions of the
Goldstone bosons $\pi^0_6,\;\pi^+_6$ and $\pi^-_6$ respectively\cite{gold}.
Therefore such interactions may give large, $CP$-violating, couplings of the
form $\eta_6\partial^{\mu}Z^0_{\mu}\partial^{\mu}Z^0_{\mu}$,
$\eta_6\partial^{\mu}W^+_{\mu}\partial^{\mu}W^-_{\mu}$,
$\eta_6\partial^{\mu}Z^0_{\mu}\partial^{\mu}W^+_{\mu}\partial^{\mu}W^-_{\mu}$
.. etc.. We consider now how these couplings can contribute to
processes at LEP and TRISTAN involving the $\eta_6$.

An essential first step is to write an effective lagrangian for the
strong (unitary gauge) longitudinal amplitudes. This will be quite different
from conventional chiral lagrangians because of the $CP$ violating amplitudes.
Indeed these amplitudes are all zero in the exact chiral limit (that is
in the absence of four-fermion $\bar{Q}Q\bar{q}q$ couplings) and so
we shall assume they are not constrained by PCAC etc.. From our
present perspective, they are simply parameters that should, presumably,
be of comparable order of magnitude. We can then add the electroweak
interaction and, in first approximation, compute to lowest order in the
electroweak couplings. For the moment we use this procedure only
implicitly to obtain some order of magnitude estimates. We shall
initially assume that, unless we argue otherwise, all momentum and mass
factors are $O(M_{Z^0})$ and effectively cancel in dimensionless
quantities. Therefore only the magnitude of electroweak couplings, small
to large mass ratios, and the order of magnitude of sextet couplings
will appear in our estimates.

First we note that, as illustrated in Fig.~3, $\eta_6 \to \gamma\gamma$ is
given by a $\partial^{\mu}W_{\mu}$ loop which, because of the unitary gauge
propagators, is clearly dominated by momenta $O(M_{Z^0})$. If we denote the
$\eta_6\partial^{\mu}W^+_{\mu}\partial^{\mu}W^-_{\mu}$ vertex by $V_1$ and
the full $\eta_6$ width by $\Gamma_{\eta_6}$, we obtain a branching ratio
\beq
\label{v1}
B_{\eta_6 \to \gamma\gamma}
\sim \alpha^2_{EM}V_1^2/m_{\eta_6}\Gamma_{\eta_6}~~~
\to~~~V_1 \sim 10^3\,\sqrt{\Gamma_{\eta_6 \to \gamma\gamma}}
\eeq
If (as we shall give arguments for below) this ratio $\sim 10^{-1}$,
and we assume $\Gamma_{\eta_6}\lsim 1$ GeV, (\ref{v1}) implies that $V_1$ is
O(1-10) on the electroweak scale.

There are contributions to $\eta_6 \to l^+l^-$ from similar loops to that of
Fig.~3 involving longitudinal $W$'s or $Z^0$'s but with one boson propagator
replaced by a lepton propagator. These amplitudes should therefore be smaller
by $O(1/M_W)$. In fact, to produce the scalar combination of helicities,
the amplitudes must involve $m_l$ and are actually $O(m_l/M_W) \sim 10^{-5}$
for an electron pair. This gives too small a coupling to allow the $\eta_6$ to
be seen at TRISTAN. A larger amplitude is obtained by producing two
photons via Fig.~3 which scatter electromagnetically into a lepton pair,
via lepton exchange. (The infra-red behavior of the photon propagators
prevents the process from vanishing as the electron mass goes to zero
and so gives an O($\aem$) amplitude rather than
O($m_e/M_{Z^0}$)\cite{con}). The resulting coupling gives a cross-section
\beq
\label{tr}
\sigma(e^+e^- \to \mbox{hadrons})~ \sim
{}~B_{\eta_6 \to \gamma\gamma}\alpha^2_{EM}
\eeq
$\sim 10\%$ of the total hadronic cross-section at TRISTAN (assuming again that
$\Gamma_{\eta_6}\lsim 1$ GeV). This is not a major effect but it is the
right order of magnitude to be compatible with the data shown in Fig.~2
and provides one argument why the two photon branching ratio of the
$\eta_6$ should be $\sim 10\%$. However, the error bars on the data
would clearly have to be significantly improved to determine that the
effect was definitively present.

In general photon emission will be strongly favored over leptons because of the
direct coupling of the photon to sextet Goldstone bosons at large momentum.
Indeed if $CP$ and $C$ are not conserved, as we are assuming, then after two
photon decay, the three photon mode could be the next most important
electroweak decay for the $\eta_6$.

Consider next the hadronic decay modes of the $\eta_6$. We anticipate that
perturbative gluon emission automatically exposes the large sextet quark
constituent mass (which is presumably $\sim 300-400$ GeV) and so is very
suppressed. Instead we expect instanton interactions to provide the major
communication between the sextet and triplet sectors. The simplest
possible final state for a decay of the $\eta_6$ mediated by an
instanton interaction would be an isotropic distribution of five quarks
and five antiquarks (one of each flavor), giving a high multiplicity
hadron state with many (mini-)jets. (There is some suggestion in the
data\cite{amy} that the increased cross-section at TRISTAN is in the higher
multiplicities). At LEP, the combination of such a state with a hard
lepton pair (i.e. $m_{l^+l^-} \sim 20-30$ GeV as in the two photon
events) could be looked for and some examples should be isolatable if
there are indeed a substantial number of such events.

At present we have no way of estimating the ratio of hadronic to two photon
branching ratios theoretically. Another phenomenological estimate, which is
clearly independent of that based on TRISTAN data, is obtained by appealing to
our suggestion\cite{ka} that hadronic diffractive production of the $\eta_6$ is
responsible for Geminion and Mini-Centauro Cosmic ray events\cite{has}, and is
also responsible (via a threshold effect) for an anomalous contribution to the
real part of the hadron elastic scattering amplitude\cite{UA4}. The number of
Geminion events thought to be associated with a 60 GeV state, suggests a
two photon cross-section of O(100-500) $\mu$bs., while the
Mini-Centauros and the threshold effect suggest a hadronic cross-section
0(2-8) mbs. Again we conclude that the two photon branching ratio should be
$\sim 10\%$.

If the LEP events are indeed produced by
$Z^0 \to \eta_6 + Z^{0\star} \to [\gamma\gamma] + [l^+l^-]$ {\em and $CP$ is
not conserved}, we can add an additional tensor vertex to (\ref{vertep}) of
the form
\beq
\label{vert}
V_{\m\n}=\Bigl[~(p.q)q_\m-q^2p_\m~\Bigr]\Bigl[~T(p,q)[(q.p)p_\n-p^2q_\n]
{}~+~L(p,q)[q^2p_\n]~\Bigr]
\eeq
where now $p$ is the momentum of the $Z^{0\star}$, so that
$T$ and $L$ are respectively invariant ``transverse'' and ``longitudinal''
amplitudes. From the above discussion we may assume that $L$ contains a large
$Z^0 \to \partial^{\mu}Z^0_{\mu}+ \g_6$ coupling (which could again be thought
of as proceeding via a $\partial^{\mu}W_{\mu}$ loop).

There is an electroweak coupling of the initial transverse $Z^0$ involved
in $L$ and so we will estimate it as
$O(gV_2)$, where g is the $SU(2)$ gauge coupling and $V_2$ is a (potentially
large) pure sextet $QCD$ amplitude. We note, however, that the longitudinal
component of a $Z^0$ propagator coupled to a lepton pair reduces to
\beq
\label{small}
\bar{u}(k)\frac{\not\!p}{M^2_{Z^0}}[\mbox{v}_l-\mbox{a}_l
\c_5]v(p-k)=-2\frac{m_l}{M^2_{Z^0}}\mbox{a}_l\bar{u}\c_5v
\eeq
where $m_l$ is the lepton mass and $u$ and $v$ are lepton spinors.
$\mbox{v}_l$ and $\mbox{a}_l$ are the vector and axial $Z^0$ couplings to the
lepton pair. The suppression factor $(m_l/M_{Z^0})$ implies there will be
{\it no neutrino pairs}, a negligible number of electron pairs, and, at
first sight, an overwhelming number of tau pairs compared to muon pairs!

Using $\alpha_W~(=g^2/4\pi)\sim 1/30$, we estimate the branching ratio
for $Z^0 \to 2\gamma + \mu^+\mu^-$ as
\beq
\label{v2}
\sim
V^2_2\alpha^2_W(m_{\mu}/M_{Z^0})^2~10^{-1}/M_{Z^0}\Gamma_{Z^0}~\sim~
V^2_2~10^{-12}~~~\to~V_2\sim 10^3~GeV
\eeq
- if (for our present purposes) we take the width of the $Z^0$ to be
$O(1)$~GeV and we estimate the LEP branching ratio to be $\sim 10^{-6}$.
Clearly (\ref{v2}) is nicely consistent with (\ref{v1}) in giving the order of
magnitude of the $QCD$ sextet quark interaction.

While the predicted absence of electron pairs is, perhaps, consistent with
the experimental situation the big question is now why there are not $\sim 300$
tau pairs for every muon pair? A perturbative correction to the tree amplitude
for tau pairs would be the loop diagram shown in Fig.~4, involving another
sextet pion vertex - $V_3$. Given that $V_3$ is $CP$-violating (and therefore
complex) we obtain a potentially negative amplitude if we take the internal
lepton line on-shell. Since this amplitude is then $O(m^2_{\tau})$ we might
suppose it to be small. However, if $Q$ is the resulting average momentum in
the loop (after the lepton line is taken on-shell) we estimate that
there is an effective perturbative expansion parameter
\beq
\label{v3}
\sim gV_3m_{\tau}/M_{Z^0}Q\sim 1
\eeq
if we take $Q\sim 10 GeV$, and assume that $V_3$ is of the same order
of magnitude as $V_1$ and $V_2$. So the effective perturbation parameter
involves the lepton mass directly and for tau pairs is sufficiently
large that the expansion breaks down. Therefore the tau pair amplitude could
well not be larger than the muon pair amplitude.

We must also discuss the production of light mass quark pairs by the
$Z^{0\star}$.
If we carry over the above analysis of leptons directly to quarks, we would
conclude that only the strange quark (with a mass of the same order of
magnitude as the muon) gives an observable cross-section which can be reliably
estimated perturbatively. However, since quark pairs carry color, they will
also interact with the initial sextet quark vertex, via $QCD$, and for
this reason alone, the amplitude can not be evaluated perturbatively.
Therefore, although we can not calculate the amplitudes, there is no
immediate conflict in the relative lack of quark (or tau) pairs.

Note that there may be a further source of lepton pairs accompanying
$\eta_6$ production. Four-fermion sextet/lepton couplings could provide a
direct (short-distance) coupling of lepton pairs into electroweak scale
instanton interactions - without going via the electroweak interaction -
and give direct $Z^0 \to \eta_6 + l^+l^-$ vertices. If $CP$ is conserved,
these amplitudes can not be large. If $CP$ is not conserved, there could
be couplings that are independent of the mass generation mechanism (involving
right-handed leptons and sextet quarks) which give large amplitudes. There are
strong constraints on such couplings which we shall not elaborate on here. We
note only that they could ultimately turn out to be necessary to understand tau
pair amplitudes. They would certainly have to play a major role if muon pairs
do not dominate over electron pairs in two photon events at LEP that are to be
explained in terms of the $\eta_6$.

In conclusion we can say that, at the order of magnitude level, a
consistent picture of the properties of the $\eta_6$ has emerged which
implies that it may indeed have been seen at both LEP and TRISTAN.

\newpage
\noindent{\bf Figure Captions}

\begin{itemize}

\item[{Fig.~1}] The two photon mass distribution for the LEP
events\cite{lep}. The bins used are 1 GeV wide and centered on the
integer values. The errors vary from experiment to experiment but are
not too different from the 0.5 GeV that our plot implies.

\item[{Fig.~2}] The error-weighted average of data from
TRISTAN\cite{trist} compared with a Standard Model prediction.

\item[{Fig.~3}] The longitudinal W loop giving the two photon decay of
the $\eta_6$.

\item[{Fig.~4}] A one-loop correction to the $\tau$-pair amplitude.

\end{itemize}

\end{document}